\documentclass[aps,prl,twocolumn,letterpaper,showpacs]{revtex4-1}
\usepackage{graphicx,amsmath,amssymb,mathrsfs,amsthm}

\begin{document}

\pagenumbering{arabic}
\title{Phonon-Induced Backscattering in Helical Edge States}
\author{Jan Carl Budich$^1$}
\author{Fabrizio Dolcini$^2$}
\author{Patrik Recher$^{1,3}$}
\author{Bj\"orn Trauzettel$^1$} 

\affiliation{$^1$Institute for Theoretical Physics and Astrophysics,
University of W$\ddot{u}$rzburg, 97074 W$\ddot{u}$rzburg, Germany;\\
$^2$Dipartimento di Scienza Applicata e Tecnologia, Politecnico di Torino, 10129 Torino, Italy;\\
$^3$Institute for Mathematical Physics, TU Braunschweig, 38106 Braunschweig, Germany}

\date\today

\begin{abstract}
A single pair of helical edge states as realized at the boundary of a quantum spin Hall insulator is known to be robust against elastic single particle backscattering as long as time reversal symmetry is preserved. However, there is no symmetry preventing inelastic backscattering as brought about by phonons in the presence of Rashba spin orbit coupling. In this letter, we show that the quantized conductivity of a single channel of helical Dirac electrons is protected even against this inelastic mechanism to leading order. We further demonstrate that this result remains valid when Coulomb interaction is included in the framework of a helical Tomonaga Luttinger liquid.
\end{abstract}
\pacs{71.10.Pm,72.15.Nj,85.75.-d}
\maketitle

During recent years, great interest has been attracted by the theoretical prediction \cite{KaneMele2005a, BHZ2006} and experimental discovery \cite{koenig2007} of the quantum spin Hall (QSH) effect. The QSH phase is a two dimensional realization of a topological insulator (TI), a class of time reversal symmetry (TRS) preserving phases that differ essentially from  trivial atomic insulators by a $\mathbb Z_2$~topological invariant \cite{kane2005b,ReXLview2010,KaneHasan}. Besides the enormous conceptual depth of topological states of matter, TI phases are also considered promising candidates as to future applications in nanoelectronics. This is due to their topologically protected transport properties which might be exploited for high precision spintronics devices operating at low power consumption. Therefore, modelling the QSH effect under experimentally relevant conditions is crucial to test the practical limitations of these protected features.

As far as the robustness of the topological protection is concerned the QSH effect is fundamentally different from the integer quantum Hall (IQH) effect. 
For the TRS breaking (IQH) insulator \cite{Klitzing1980,Laughlin1981,TKNN1982} topological order leads to a quantization of conductivity to impressive accuracy. 
In the IQH regime, edge state transport is chiral, meaning that the density of states for subgap backscattering vanishes which excludes such processes by simple phase space arguments.
In contrast, in the TRS preserving QSH phase a single pair of helical edge states induced by bulk boundary correspondence is supported at the edge of the QSH bar. This means that both right- and leftmovers exist at a single edge. However, states of opposite direction of motion are Kramers partners due to TRS. The well known topological protection of a single pair of helical edge states against backscattering in this scenario can be mathematically illustrated by the following simple argument \cite{XuMoore06}. Let $\lvert \phi \rangle$~and $\lvert \psi\rangle =T\lvert \phi\rangle$~be Kramers partners. Then as long as $H$~is a TRS preserving Hamiltonian
\begin{align}
\langle \psi\rvert H\lvert \phi \rangle =& \langle \phi\rvert H\lvert \psi \rangle^* = \langle T\phi\rvert TH\lvert \psi \rangle=\nonumber\\
&\langle\psi\rvert H T\lvert \psi \rangle = \langle\psi\rvert H T^2\lvert \phi \rangle =-\langle\psi\rvert H \lvert \phi \rangle
\label{eqn:topprotect}
\end{align}
i.e. the matrix element for scattering between the Kramers partners vanishes.
Note that argument (\ref{eqn:topprotect}) of protection relies on two fundamental constraints: First, only single electron processes are considered. Second, since Kramers partners are degenerate states, it only precludes elastic backscattering. 
Within the validity of these restrictions extensive studies of the helical Tomonaga Luttinger liquid (hTLL) \cite{XuMoore06,Wu2006} representing a single pair of helical edge states have shown that Anderson localization is avoided \cite{KaneMele2005a} in the presence of TRS preserving disorder and that TRS breaking magnetic impurities can open a gap in these systems \cite{Maciejko2009}. Furthermore, interedge backscattering can occur if the QSH sample is locally narrowed down to a quantum point contact \cite{Stroem2009,teo2009, HouChamon2009, CX2011,Schmidt2011} or if two QSH bars are brought close to each other \cite{NagaosaTwoEdge}.
In general, backscattering at a single helical edge requires spin flip processes. In realistic setups, these are induced by Rashba spin orbit coupling (SOC) originating from unavoidable potential fluctuations. Preserving TRS, Rashba SOC cannot cause \emph{single electron} elastic backscattering, though. 
However, relaxing the single electron processes constraint by additionally including Coulomb interaction, two electron backscattering processes have been shown to arise in these systems \cite{Stroem2010b}. 
Such backscattering terms are well known to be allowed by TRS \cite{Wu2006}.

\begin{figure}
\centering
\includegraphics[width=170pt]{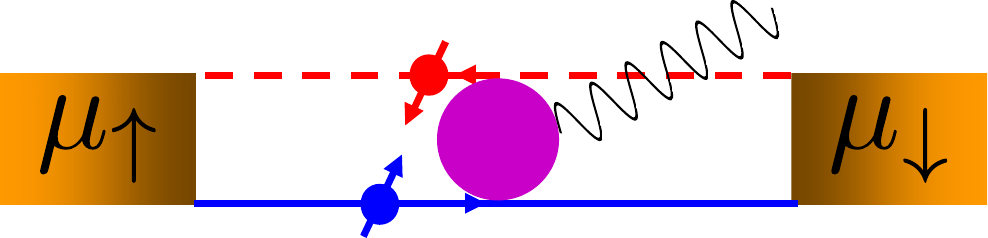}
\caption{Pair of helical edge states with two terminals and a Rashba impurity coupling the two channels. The wavy line illustrates the presence of electron phonon coupling in the system.}
\label{fig:helicalRashba}
\end{figure}

 Under realistic experimental conditions finite temperature and bias voltage also imply the presence of phonons, i.e. inelastic processes that undermine the second constraint for the validity of argument (\ref{eqn:topprotect}). It is thus of crucial importance to investigate the influence of this dissipative mechanism on the topological protection. Here, we show two important results of the helical edge states in presence of two TRS preserving perturbations: Rashba SOC and electron phonon coupling. First, we demonstrate that in this scenario there is no strict protection against inelastic \emph{single electron} backscattering. Second, we find that for helical Dirac fermions the leading order contribution of this mechanism vanishes, supporting the protection for practical purposes. We further demonstrate how this additional robustness fully survives in presence of Coulomb interaction, i.e. in a hTLL with electron phonon coupling and Rashba SOC. In a nonequilibrium transport calculation for the hTLL, we take the electron phonon coupling into account exactly by integrating out the phonons using a Keldysh contour path integral representation of the generating functional. Our analysis is relevant for any realization of the hTLL as a one dimensional system.
{\it Model without Coulomb interaction---}   
We investigate a single pair of helical Dirac fermions coupled linearly to longitudinal acoustic phonons. The two species of electrons are coupled via Rashba SOC (see Fig. \ref{fig:helicalRashba}). In most parts of this work we will have a sharp impurity-like scattering potential in mind which brings about momentum transfer on the order of $2 k_F$, where $k_F$~is the Fermi wave vector.
We represent the helical fermionic fields as a spinor $\Psi = (\Psi_{R\uparrow},\Psi_{L\downarrow})^T\equiv(\Psi_+,\Psi_-)^T$.  The free electron Hamiltonian then reads
\begin{align*}
H_{\rm hl}=\int dx \Psi^\dag(x) p\sigma_z \Psi(x)
\end{align*}
where $p=-i\partial_x$~is the momentum operator an $\sigma_z$~is a Pauli matrix in spin space. The two most relevant Rashba-terms induced by a spatially dependent electric field in $z$-direction are given by \cite{Stroem2010b,Dietrich2010}
\begin{align*}
H_R=\frac{1}{2}\int dx\Psi^\dag(x)\left(\left\{\alpha_1(x), p\right\}+\left\{\alpha_3(x), p^3\right\}\right)\sigma_y \Psi(x)
\end{align*}
Note that only odd powers of $p$~are allowed by TRS.
Electron phonon coupling to the displacement field $d$~of longitudinal acoustic phonons is modelled by the Hamiltonian \cite{LossMartin}
\begin{align*}
H_{\rm ep}= \lambda\int dx \Psi^\dag(x)\sigma_0\Psi (x)\partial_xd(x)
\end{align*}
with all dimensionful constants absorbed into $\lambda$. The free phonon dynamics is governed by
\begin{align*}
&H_p=\frac{1}{2}\int dx\left[(\Pi_d(x))^2+c^2(\partial_x d(x))^2\right]
\end{align*}
where $c$~is the acoustic phonon velocity in units of the electronic Fermi velocity and $\Pi_d$~is the conjugate momentum of $d$. We model the phonons for a strictly 1D system which corresponds to an in transverse direction perfectly localized edge state. Later on, we will see that our key results do not critically depend on the details of the phonon model.
The total Hamiltonian of our setup is then given by
\begin{align}
H=H_{\rm hl}+H_p+H_R+H_{\rm ep}=H_0+H_I
\end{align}
where $H_0= H_{\rm hl}+H_p$~is the free Hamiltonian whereas $H_I=H_R+H_{\rm ep}$~encompasses the coupling terms.

\begin{figure}[htp]
\centering
\includegraphics[width=120pt]{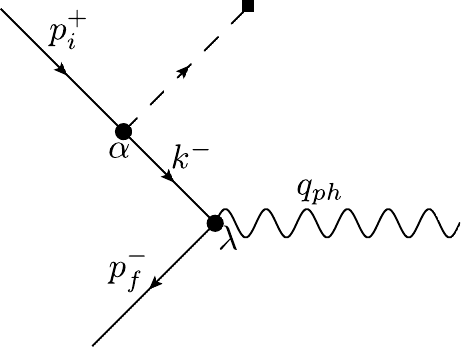}
\includegraphics[width=120pt]{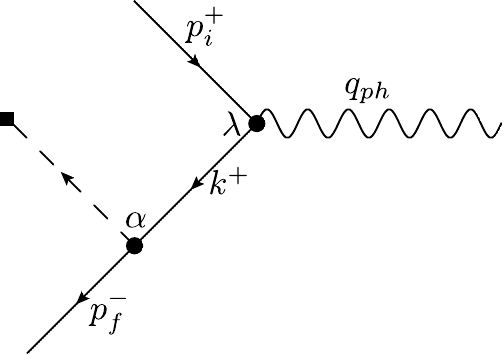}
\caption{Tree diagrams for lowest order backscattering. Dashed line with square represents the external Rashba potential. Wavy line denotes the phonon propagator.}
\label{fig:diagram}
\end{figure}

\begin{figure}
\centering
\includegraphics[width=110pt]{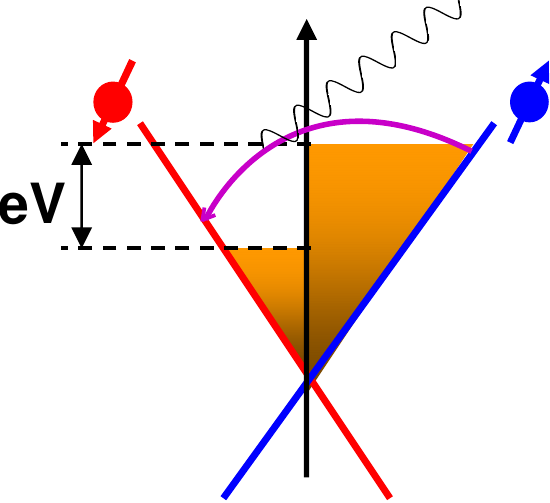}
\caption{Voltage configuration to pass a spin up current from the left to the right. The bias $V$~opens an energy window for inelastic phonon scattering.}
\label{fig:helicalBias}
\end{figure}

{\it Inelastic backscattering---}
We will now demonstrate how the combination of $H_R$~and $H_{\rm ep}$~will in principle be able to cause \emph{single electron} backscattering at a single edge of a QSH insulator. As observed above, since $H_R$~is TRS preserving it cannot cause elastic single electron backscattering. Due to its offdiagonal structure in spin space it couples opposite spins though. In contrast, $H_{\rm ep}$~does not mix different spin species but can bring about energy dissipation by virtue of energy transfer from the electronic degrees of freedom to phonons. Thus the second order in $H_I$~diagrams shown in Figure \ref{fig:diagram} which are first order in $H_{\rm ep}$~and in $H_R$~cause backscattering at finite bias (see Fig. \ref{fig:helicalBias}) if their contribution does not vanish for momentum transfer $p_i-p_f\approx 2k_F$. 
We consider scattering between a right mover $\lvert p_i^+\rangle$~and a left mover $\lvert p_f^-,q_{\rm ph}\rangle$~with an additional phonon. Up to second order the corresponding scattering matrix element $M_{\rm if}$~can be written as
\begin{align*}
M_{\rm if}= \langle p_f^-,q_{\rm ph} \rvert H_IG_0H_I\lvert p_i^+\rangle
\end{align*}
where $G_0$~is the free propagator corresponding to $H_0$. Interestingly, the lowest order contribution of the Rashba term linear in momentum associated with $\alpha_1$~vanishes due to a nontrivial destructive quantum interference of the two contributing diagrams which stems from the linearity of both $H_{\rm hl}$~and the $\alpha_1$-Rashba term. To show that this is not due to any fundamental symmetry like TRS we calculate the same matrix element for the Rashba term associated with $\alpha_3$~which yields
\begin{align*}
\lvert M_{\rm if}\rvert^2=\frac{\lambda^2 c}{16\pi}\tilde \alpha^2_3\left(q_{\rm ph}+p_f^--p_i^+\right)\lvert q_{\rm ph}\rvert^5
\end{align*}
where $\tilde\alpha_3(k)$~is the Fourier transform of $\alpha_3(x)$. An analytical Fermi's golden rule calculation at zero temperature for a $\delta$-shaped Rashba impurity $\alpha_3(x)=\alpha_3\delta(x)$ yields a backscattering current
\begin{align}
I_{\rm BS}= \frac{\alpha_3^2\lambda^2 e}{672 \pi^2 c^5}V^7.
\label{eqn:ibs}
\end{align}
This contribution will be negligible at low bias $V$~reflecting the irrelevance of the $p^3$-Rashba term. The importance of this nonvanishing result is, that it demonstrates how in principle inelastic \emph{single electron} backscattering can occur at finite bias even though the perturbations Rashba SOC and electron phonon coupling preserve TRS.  The lowest order nonvanishing matrix element for the $\alpha_1$~term could be third order in $\alpha_1$~which has the same relevance as $\alpha_3$~in renormalization group (RG) sense. Also quadratic corrections to the linear electronic dispersion which might become relevant at higher energies can give rise to $\alpha_1$~backscattering.
We conclude that for the system without Coulomb interaction the most relevant inelastic backscattering contributions allowed by TRS and phase space arguments cancel. This can be interpreted as an enhanced precision of the quantized conductivity of the helical edge states at finite temperature/bias going beyond the topological protection only pertaining to elastic scattering.

{\it hTLL with Coulomb interaction ---}
Now, we want to investigate whether the observed robustness of the helical Dirac fermions against inelastic backscattering by virtue of the $\alpha_1$-Rashba SOC term, i.e. the most relevant TRS preserving term coupling opposite spins, persists in the presence of Coulomb interaction.
In order to account for Coulomb interaction we represent the fermionic degrees of freedom in terms of a hTLL using the bosonization identity
\begin{align}
\psi_{\pm}= \frac{1}{\sqrt{2\pi a}}\eta^{\pm}\text{e}^{\mp i\sqrt{\pi}(\varphi\pm \theta)},
\end{align}
where $\psi_{\pm}$~now denote the slowly varying fields with a factor of $\text{e}^{\pm ik_Fx}$~separated off, $a$~is the high energy cutoff of the model and $\varphi,~\theta$~are the bosonic phase field and its dual respectively. In the thermodynamic limit the Klein factors $\eta^{\pm}$~obey the algebra of Majorana fermions. Absorbing a prefactor of $\frac{1}{\pi}$~by redefining $\lambda$, the electron phonon coupling can be represented as
\begin{align*}
H_{\rm ep}=\lambda\int dx \partial_x\varphi(x)~\partial_x d(x).
\end{align*}
The free hTLL Hamiltonian reads
\begin{align*}
H_{\rm hTLL}=\frac{1}{2}\int dx \left[\Pi^2_\varphi(x)+\frac{1}{g^2}\left(\partial_x \varphi(x)\right)^2\right]
\end{align*}
with the interaction strength parameter $g<1$~modelling repulsive Coulomb interaction. The $\alpha_1$-Rashba Hamiltonian with $\alpha_1(x)=\alpha\delta(x)$~in bosonized form yields
\begin{align*}
H_R=\frac{i\alpha}{\sqrt{\pi} a}\eta^+\eta^-\left.:\left(\partial_x\theta(x)\right)\cos\left(\sqrt{4\pi}\varphi(x)\right):\right|_{x=0}
\end{align*}
where the dots denote normal ordering.

We want to calculate the average current $I(x,t)=\frac{e}{\sqrt{\pi}}\partial_t\langle\varphi(x,t)\rangle$~due to an applied bias V. Such expectation values can be most easily represented for practical calculations in terms of the generating functional
\begin{align}
Z\left[J\right]=\int D(\varphi,\theta,d)\text{e}^{iS_0-i\int_cH_R+i\frac{e}{\sqrt{\pi}}E^T\sigma_3\varphi+\frac{i}{\sqrt{2}}J^T\varphi},
\label{eqn:zofj}
\end{align}
where $S_0$~encompasses the electron phonon system without the Rashba impurity, $\int_c$~is along the Keldysh contour, $\sigma_3$~is a Pauli matrix in Keldysh space, and scalar products like $J^T\varphi$~involve an integration over real space and time. The applied bias is modelled by $E(x,t)$~along the lines of Ref.~\cite{Dolcini2005}.

To make further analytical progress we now integrate out the phonons on the Keldysh contour.
The part of the Lagrangian involving the phonon field $d$ reads
\begin{align*}
L_d = \frac{1}{2}\left(\left(\partial_t d\right)^2-c^2 (\partial_x d)^2\right)-\lambda (\partial_x\varphi) (\partial_xd)
\end{align*}
The phonon dependent part of the action can be represented on the Keldysh contour as
\begin{align*}
S_d = \frac{1}{2}d^TG_{\rm ph}^{-1}d+\lambda d^T \sigma^3 \partial_x^2\varphi.
\end{align*}
with the phonon propagator $G_{\rm ph}$. Performing the Gaussian integral
\begin{align*}
\int Dd~ \text{e}^{iS_d(d,\varphi)}=\text{e}^{iS_{\text{diss}}(\varphi)}
\end{align*}
in the rotated Keldysh basis
\begin{align*}
\begin{pmatrix} \varphi^+ \\ \varphi^- \end{pmatrix}\rightarrow \frac{1}{\sqrt{2}}\begin{pmatrix} 1 & 1 \\ 1 & {-1} \end{pmatrix}\begin{pmatrix} \varphi^+ \\ \varphi^- \end{pmatrix}\equiv U\varphi
\end{align*}
yields for the dissipative action
\begin{align}
S_{\text{diss}}=-\frac{\lambda^2}{2}(\partial_x^2\varphi)^T\sigma_1 G_{\rm ph}\sigma_1 \partial_x^2 \varphi,
\label{eqn:sdiss}
\end{align}
where $\sigma_1 G_{\rm ph} \sigma_1=\begin{pmatrix} 0 & G_{\rm ph}^A \\ G_{\rm ph}^R & G_{\rm ph}^K \end{pmatrix}$.
By this dissipative action the inverse free electron Green function in Fourier space is changed to the following dressed version
\begin{align}
\left(G_e^{-1}(k,\omega)\right)_{\varphi\varphi}\rightarrow  \left(G_e^{-1}(k,\omega)\right)_{\varphi\varphi}-\frac{\lambda^2 k^4}{\omega^2-c^2k^2}
\label{eqn:Greendress}
\end{align}
This result generalizes to the Keldysh formalism a similar imaginary time calculation carried out in Ref.~\cite{LossMartin}. The retarded, advanced and Keldysh part of this Green function can be calculated exactly.
From now on the free action $S_0(\varphi,\theta)$~refers to the effective action where the phonons have been integrated out.
To calculate the current we basically have to evaluate
\begin{align}
\langle \varphi(x)\rangle=\frac{-i}{\sqrt{2}}\left.\frac{\delta Z[J]}{\delta J(r)}\right|_{J=0}
\label{eqn:phiexp}
\end{align}
which can be done along the lines of Ref.\cite{Dolcini2005}. The equation of motion $\partial_t \varphi =-\partial_x\theta~$ inside the free correlators averaged with $S_0(\varphi,\theta)$~remains valid for the free action which is dressed by the phonon dissipation. That is because $S_{\text{diss}}$ (see Eq. (\ref{eqn:sdiss})) depends only on $\varphi$~and not on its conjugate momentum $\partial_x\theta$. Using only this property of the electron phonon coupling the current to second order in $\alpha_1$~ is readily shown to be zero proving that our result for the case without Coulomb interaction persists in the nonequilibrium hTLL.

This result can be understood on more general grounds. The fact that $S_0$~remains quadratic in $\Pi_\varphi=\partial_x\theta$~means that the following argument first brought forward in Ref.~\cite{Stroem2010b} for a hTLL in equilibrium without electron phonon interaction can be used for our setup as well: Integrating out $\Pi_\varphi$~ in the path integral representation of the generating functional will produce terms proportional to $(\partial_t\varphi)^2,~\alpha(\partial_t\varphi)\cos(\sqrt{4\pi}\varphi),\alpha^2\cos(\sqrt{4\pi}\varphi)^2$~in the action. While the first contribution is the term  well known from the free hTLL case, the second one is a pure gauge which can be dropped. The third term can up to a constant be written as $\cos(\sqrt{16 \pi}\varphi)$~which is a two electron TRS preserving backscattering term \cite{Wu2006}. Note that the presence of the source term $J^T\varphi$~ in Eq. (\ref{eqn:zofj}) does not affect this argument. Thus we have found that the $\alpha_1$-Rashba term cannot lead to a single electron backscattering term in presence of any external spin independent dissipation which couples linearly to the electron density. We again point out that this result goes beyond the topological protection of the hTLL. It is due to the quadratic form of $S_0$~in $\Pi_\varphi$, the Luttinger liquid analogue of the linear dispersion of helical Dirac fermions on which our result without Coulomb interaction relied. Bosonizing the $\alpha_3$-Rashba term implies terms up to third power in $\partial_x \theta$~thus breaking the quadratic form of $S_0$~in $\Pi_\varphi$. Furthermore mixed terms like $(\partial_x \varphi)(\partial_x\theta)^2$~will occur which render the modifications of free Green function (see Eq. (\ref{eqn:Greendress})) by the presence of phonons important. These observations are perfectly compatible with our nonvanishing result for single electron backscattering in the presence of $\alpha_3$~(see Eq.(\ref{eqn:ibs})).

In summary, we have studied helical Dirac fermions in the presence of electron phonon coupling and Rashba SOC, which preserve TRS. We have shown that, although TRS does not provide a protection against inelastic scattering, the current carried by the helical states in presence of a finite bias is not changed to leading order. Furthermore, we have proven that this result still holds for a hTLL including Coulomb interaction. The linear dispersion of the helical edge states of a QSH bar has been nicely verified experimentally \cite{koenig2007} and is an exact feature of the four band model for inverted HgTe/CdTe quantum wells introduced in Ref. \cite{BHZ2006}. Therefore, our analysis is not only interesting for the abstract model of a hTLL. It supports the robustness of the quantized subgap conductance of a QSH sample beyond the well known argument (\ref{eqn:topprotect}) of topological protection. Our results turn out to be not restricted to the coupling to longitudinal acoustic 1D phonons. In fact, we have shown that any external bath coupling linearly to the electron density cannot give rise to inelastic single electron backscattering in presence of linear in $k$~spin orbit coupling. In HgTe/CdTe layer structures external coupling mechanisms, e.g coupling to charge puddles in the bulk, are likely to cause phase decoherence which gives rise to additional backscattering. However, such effects are not intrinsic features of the hTLL and can in principle be contained by improving the sample quality. In contrast, electron phonon coupling and Coulomb interaction are intrinsic mechanisms the role of which we have investigated for a generic realization of the hTLL. 

We acknowledge interesting discussions with Markus Kindermann and financial support by the DFG-JST Research Unit "Topotronics" (JCB and BT), the Emmy Noether program (PR), and the Vigoni program (FD and BT).


\end{document}